\documentclass[journal=jpccck,manuscript=article]{achemso} 

\usepackage{chemformula}
\usepackage[T1]{fontenc}

\usepackage[version=3]{mhchem} 
\usepackage{indentfirst}

\author{Mingda Ding}
\affiliation[OUKobaYoshi]
{Department of Applied Physics, Graduate School of Engineering, The University of Osaka, Suita, Osaka 565-0871, Japan}
\email{ding@ap.eng.osaka-u.ac.jp}

\author{Nanami Sakurai}
\affiliation[OUKobaYoshi]
{Department of Applied Physics, Graduate School of Engineering, The University of Osaka, Suita, Osaka 565-0871, Japan}

\author{Man Shen}
\affiliation[OUKobaYoshi]
{Department of Applied Physics, Graduate School of Engineering, The University of Osaka, Suita, Osaka 565-0871, Japan}

\author{Taiki Inoue}
\affiliation[OUKobaYoshi]
{Department of Applied Physics, Graduate School of Engineering, The University of Osaka, Suita, Osaka 565-0871, Japan}
\email{inoue.taiki@ap.eng.osaka-u.ac.jp}

\author{Takahisa Matsuzaki}
\affiliation[OUKobaYoshi]
{Department of Applied Physics, Graduate School of Engineering, The University of Osaka, Suita, Osaka 565-0871, Japan}
\alsoaffiliation[OUMatsuzaki]
{Division of Precision Engineering and Applied Physics, Center for Future Innovation, Graduate School of Engineering, The University of Osaka, Suita, Osaka 565-0871, Japan}

\author{Takuo Mizuno}
\affiliation[OUKobaYoshi]
{Department of Applied Physics, Graduate School of Engineering, The University of Osaka, Suita, Osaka 565-0871, Japan}

\author{Kento Suzuki}
\affiliation[OUKobaYoshi]
{Department of Applied Physics, Graduate School of Engineering, The University of Osaka, Suita, Osaka 565-0871, Japan}

\author{John Isaac Enriquez}
\affiliation[OUMori]
{Department of Precision Engineering, Graduate School of Engineering, The University of Osaka, Suita, Osaka 565-0871, Japan}

\author{Harry Handoko Halim}
\affiliation[OUMori]
{Department of Precision Engineering, Graduate School of Engineering, The University of Osaka, Suita, Osaka 565-0871, Japan}

\author{Yuji Hamamoto}
\affiliation[OUMori]
{Department of Precision Engineering, Graduate School of Engineering, The University of Osaka, Suita, Osaka 565-0871, Japan}

\author{Yuta Nishina}
\affiliation[Okayama]
{Research Institute for Interdisciplinary Science, Okayama University, Okayama, Okayama 700-8530, Japan}

\author{Hiroshi Y. Yoshikawa}
\affiliation[OUKobaYoshi]
{Department of Applied Physics, Graduate School of Engineering, The University of Osaka, Suita, Osaka 565-0871, Japan}

\author{Yoshitada Morikawa}
\affiliation[OUMori]
{Department of Precision Engineering, Graduate School of Engineering, The University of Osaka, Suita, Osaka 565-0871, Japan}

\author{Yoshihiro Kobayashi}
\affiliation[OUKobaYoshi]
{Department of Applied Physics, Graduate School of Engineering, The University of Osaka, Suita, Osaka 565-0871, Japan}
\email{kobayashi@ap.eng.osaka-u.ac.jp}

\title{Mechanical bistability and hysteresis in graphene--CNT hybrid systems: from atomistic simulations to macroscale structural responses}


\begin{document}

\begin{abstract}
Hybrid systems composed of graphene (Gr) and carbon nanotubes (CNTs), such as films and aerogels, have attracted broad attention for applications in electronics, mechanics, energy, and environmental science. Since the microstructures of Gr--CNT hybrids strongly affect their properties, it is essential to establish mechanical principles that govern these structures. In this study, we investigated the structural stability and mechanical behavior of Gr--CNT hybrid systems by combining molecular dynamics (MD) simulations and nanoindentation experiments. MD simulations of stacked Gr--CNT structures, in which two Gr layers confine CNTs between them, identified the energetically stable configurations and their governing parameters, i.e., intertube spacing, CNT diameter, and wall number. Specifically, under certain conditions, the structures exhibit mechanical bistability with two stable configurations: adhesion and separation of the Gr layers, arising from the competition between interlayer van der Waals attraction and elastic deformation of Gr and CNTs. Simulated loading--unloading curves display hysteresis and energy dissipation related to the stable configurations. In addition, reduced graphene oxide (rGO)--CNT hybrid films were experimentally fabricated as macroscopic assemblies of the unit structures modeled in the simulations. Atomic force microscopy-based nanoindentation measurements on the rGO--CNT films exhibit clear hysteresis and higher dissipation energy compared with pure rGO, in good agreement with the simulation results. These results provide valuable insights into Gr--CNT hybrid systems and offer guidance for designing microstructures with enhanced properties for advanced applications. 
\end{abstract}

\section{Introduction}

Hybrid systems of graphene (Gr)\cite{Novoselov2004, Novoselov2005} and carbon nanotubes (CNTs)\cite{Iijima1991, Iijima1993}, including composite films, sponges, and aerogels\cite{nardecchia2013three}, have been extensively studied for their potential in various advanced applications \cite{kim_graphene_2012,afroze_hierarchical_2021,liu_lightweight_2022,shang_interfacial_2022,kholmanov_optical_2015,liu_planar_2015,liao_enhanced_2019}. Gr is a revolutionary material characterized by its atomic thickness, exceptional electrical and thermal conductivity, and remarkable mechanical stiffness \cite{geim2009graphene}. The incorporation of CNTs into Gr films\cite{nardecchia2013three} can prevent Gr layers from restacking and offers a cross-linking effect \cite{bao_vacancy-induced_2025}. As a result, Gr--CNT hybrid systems exhibit enhanced mechanical robustness, porosity, electrical conductiviy, and superelasticity \cite{kim_graphene_2012,mu_graphene-carbon_2015,li_fabrication_2017,afroze_hierarchical_2021,shang_interfacial_2022}. Owing to these combined properties, Gr--CNT hybrid systems have been applied in diverse functional materials and devices. Specifically, their porous structure with a large surface area provides an ideal scaffold for adsorbent materials in catalytic processes and water purification\cite{tristan-lopez_large_2013,wan_graphenecarbon_2016,yin_application_2020}. Their excellent thermal and electrical conductivity further broadens their applications to supercapacitors\cite{mu_graphene-carbon_2015,yu_scalable_2014,wang_direct_2020}, batteries\cite{zhang2017high, peng_electrochemical_2020}, and thermal management\cite{lv_super-elastic_2016,du_wide_2021}. In addition, the flexibility and lightweight nature of Gr--CNT hybrid systems render them suitable for use in strain sensors\cite{li_fabrication_2017,liu_lightweight_2022,afroze_hierarchical_2021,shang_interfacial_2022}, artificial muscles\cite{yocham_mechanical_2018}, and dampers\cite{kim_graphene_2012}.

The microstructure of Gr--CNT hybrid systems strongly governs their overall characteristics\cite{nardecchia2013three}, ranging from porosity and flexibility to electronic structure and transport behavior. The adhesion between single Gr layers causes their restacking, forming multilayer structures with strong interlayer coupling\cite{Latil2007}. Incorporating nanomaterials such as CNTs as nanospacers helps maintain the Gr layer separation\cite{shang_interfacial_2022, Xu2022, ding_experimental_2024}; however the single-atom thickness of Gr results in low bending stiffness. When the fraction of CNT nanospacers is insufficient, the Gr layers may still bend and adhere, leading to undesirable effects such as stacking-induced transport degradation\cite{Sui2009} and a reduced surface-to-volume ratio\cite{zhang2017high}. To design stable hybrid structures, it is crucial to clarify the role of nanoscale adhesion and deformation of Gr and CNTs in determining the resulting microstructures. Understanding these interactions would not only provide fundamental insight into van der Waals (vdW) interfaces but also offer practical guidance for developing lightweight, flexible, and conductive hybrid films for device applications\cite{afroze_hierarchical_2021,liu_lightweight_2022,shang_interfacial_2022,kholmanov_optical_2015,liu_planar_2015,liao_enhanced_2019}.

In previous studies, molecular dynamics (MD) simulations have been employed to investigate the mechanical properties of multilayer Gr--CNT stacked structures \cite{menon_mechanical_2020,xu_configuration_2023,bao_vacancy-induced_2025}. The models were simplified from three-dimensional (3D) sponges of Gr to stacking structures to reduce the computational cost. This allowed studying the elastic constants of the multilayer Gr--CNT stacked structures \cite{menon_mechanical_2020} and the deformation of a CNT encapsulated by two Gr layers \cite{xu_configuration_2023}. However, the deformation of this structure under external forces has not been sufficiently explored. The coarse-grained method was also implemented to study the fracture behavior of 3D sponges under pressure loading\cite{wang_mechanical_2023}, but further research on the unloading process associated with such structures is required. Cyclic loading--unloading is essential for the reusability of such materials in practical applications. Although the hysteresis of the loading--unloading curve has been commonly observed in previous experiments\cite{kim_graphene_2012,li_fabrication_2017,afroze_hierarchical_2021,zhao_mechanical_2021}, the underlying mechanism has not been theoretically elucidated. In our previous work, a Gr--CNT hybrid system supported on a substrate was studied using a combination of experiments and MD simulations\cite{ding_reduction_2023}. However, the loading response and multistable characteristics remain unexplored. Additionally, the boundary conditions in that study were set to simulate substrate supported structures, rather than macroscopic Gr--CNT hybrids such as sponges, in which the Gr layers and CNTs are nearly suspended.

In this study, we systematically investigated the microstructure and mechanical behavior of Gr--CNT stacked structures using MD simulations and experiments. The MD simulations reveal the presence of mechanical bistability with two energetically stable states corresponding to adhesion and separation configurations, which depend on the CNT diameter, intertube spacing, and wall number. An energy analysis elucidates the physical origin of the bistability. The simulated loading--unloading responses exhibit pronounced hysteresis and energy dissipation, which originate from transitions between the two stable configurations. Finally, nanoindentation experiments on macroscopic Gr--CNT hybrid films confirm the hysteretic energy dissipation behavior, establishing a direct connection between nanoscale structural stability and macroscopic mechanical response.

\section{Methods}
\subsection{MD simulations}
\begin{figure*}[!htbp]
	\includegraphics[width=0.8\textwidth]{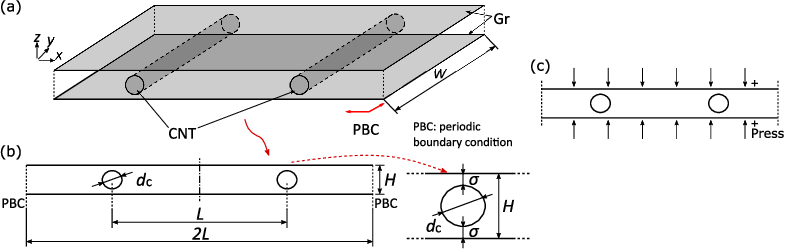}
	\centering
	\caption{Schematics of MD simulation system in (a) three-dimensional view and (b) front view. The system contains two Gr layers and two CNTs aligned in parallel with the $y$ direction. The width of the Gr in the $y$ direction is $W$, equal to the length of CNTs. The length of Gr in the $x$ direction is $2L$. Systems with different intertube distances $L$ and CNT diameters $d_{\rm c}$ were simulated. The initial distance between the two Gr layers is $H = d_{\rm c} + 2\sigma$, as shown in the enlarged view in (b). PBCs were applied in the in-plane $x$ and $y$ directions. (c) Schematic of pressure loading applied to the Gr layers. The two sets of arrows represent opposing forces acting on the Gr layers.}
	\label{fig:MDsystem}
\end{figure*}

In the simulation model, two CNTs are sandwiched between two Gr layers (Figure \ref{fig:MDsystem}). The CNTs are aligned parallel to the $y$ direction with an intertube distance of $L$. The Gr layers have a length of $2L$ along the $x$ direction and a width of $W=6.7$ nm along $y$ direction. Initially, the interlayer distance between the two Gr layers was set to $H=d_{c}+2\sigma$, where $d_{c}$ is the CNT diameter and $\sigma$ is the interlayer distance between Gr layers in graphite. The periodic boundary condition (PBC) was applied along the in-plane $x$ and $y$ directions of the simulation box for the model to simulate the CNT array aligned at equal distances of infinite length. Unlike the substrate-supported boundary condition used in our previous study\cite{ding_reduction_2023}, both Gr layers were modeled as free-standing sheets so as to capture the quasi-suspended nature of the Gr--CNT units in macroscopic hybrids. The adaptive intermolecular reactive empirical bond order (AIREBO) potential\cite{stuart_reactive_2000} was used to describe the short-range interactions between carbon atoms. The long-range vdW interactions between the Gr layers and CNTs were described using the Lennard--Jones (LJ) potential as follows: 
\begin{equation}
E_\text{LJ}=4\epsilon \left( \frac{\sigma ^{12}}{r^{12}}-\frac{\sigma ^6}{r^6} \right), 
\label{eqn:LJ12_6}
\end{equation}
where $r$ is the distance between two atoms, $\epsilon$ is the depth of the potential well, and $\sigma$ represents the distance when the LJ energy $E_\text{LJ}$ is zero. The minimization point $E_\text{LJ}=-\epsilon$ appears at $r=\sqrt[6]{2}\sigma$, and $E_\text{LJ}\rightarrow 0$ when $r\rightarrow \infty$. The equilibrium distance between Gr layers in graphite also equals to $\sigma$. The parameters were set to $\epsilon= 0.00284 \ \text{eV}$ and $\sigma= 0.34 \ \text{nm}$\cite{stuart_reactive_2000}. In MD simulations, a cutoff radius $r_\text{cutoff}$ must be defined for long-range interactions to reduce the computational cost. To induce vdW interactions between the upper and lower Gr layers, $r_\text{cutoff}$ was set to be larger than the interlayer distance ($r_\text{cutoff} > H$). The system was relaxed in the NVT ensemble with a temperature of 300 K and a time step of 1 fs. The energy and configuration of the system were output every 1000 steps in the relaxation process to obtain the energetically stable configuration. All MD simulations were performed using the open-source software, large-scale atomic/molecular massively parallel simulator (LAMMPS)\cite{thompson_lammps_2022}. 

The compressive behavior of the Gr--CNT stacked structure was investigated by applying a loading--unloading process. To simulate a pressure load on the structure, a uniform force $F$ was applied to each atom in the Gr layers (Figure \ref{fig:MDsystem}c). The pressure is given by $P=F/S_0$, where $S_0=0.026 \ \rm nm^2$ is the planar area of a carbon atom in a Gr layer. The loading process was conducted incrementally, with $F$ increasing and decreasing in steps of 0.0016 nN. After each step, the system was relaxed for 100 ps to obtain a stable structure.

\subsection{Fabrication and characterization of Gr--CNT hybrid films}
\begin{figure*}[!htbp]
	\includegraphics[width=0.95\textwidth]{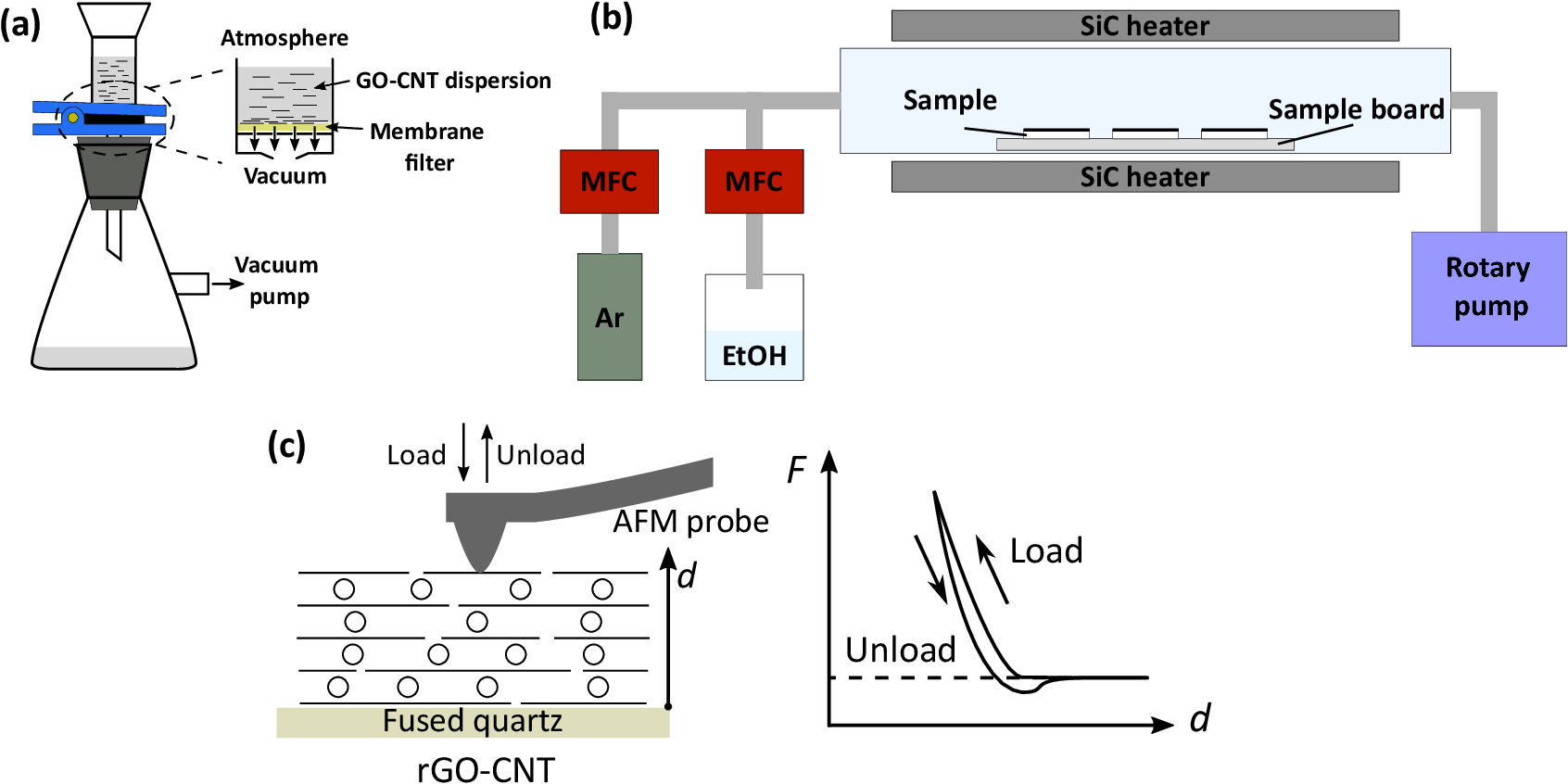}
	\centering
	\caption{Schematics of experimental processes for sample fabrication and nanoindentation measurements. (a) Fabrication of GO films and GO--CNT hybrid films via vacuum filtration. (b) Reduction process to obtain rGO and rGO--CNT films using a high-temperature electric furnace with a mixed gas of ethanol (EtOH) and Ar. (c) Nanoindentation loading--unloading measurement performed on the films by controlling the vertical movement of an AFM probe. The corresponding plot of loading force versus probe distance was obtained accordingly.}
	\label{fig:ExpMethod}
\end{figure*}

\textbf{Sample preparation} The mechanical behavior of macroscopic Gr--CNT stacked structures was experimentally investigated using the nanoindentation method. A process using graphene oxide (GO) was employed to fabricate the Gr-based hybrid system, because it enables scalable fabrication of macroscopic assemblies\cite{stankovich2006graphene,stankovich2007synthesis} and has been adopted for diverse applications of Gr-based hybrids\cite{afroze_hierarchical_2021,liu_lightweight_2022,shang_interfacial_2022, xiong2015ultrarobust,zhao2018reduced,xu2023improved}. Multilayer reduced graphene oxide (rGO) films with and without CNTs were fabricated from GO suspensions via a combination of vacuum filtration and high-temperature reduction. A 0.1 wt\% aqueous GO suspension obtained using a modified Hummers method\cite{morimoto2016tailoring} and a 0.1 wt\% aqueous CNT suspension (ZEONANO 02DS-WA-RD, average tube diameter of 4 nm)\cite{hata2004water} were prepared. The GO and CNT suspensions were mixed at a mass ratio of 16:1 to obtain a GO--CNT suspension. Subsequently, GO and GO--CNT films were fabricated through filtration of the respective suspensions (Figure \ref{fig:ExpMethod}a). After removing the filter membrane with acetone, the films were transferred onto fused quartz substrates. Reduction of the GO and GO--CNT films to obtain rGO and rGO--CNT films, respectively, was performed using a high-temperature reduction method with ethanol as the reducing agent \cite{ishida_anomalous_2016,Xu2021} (Figure \ref{fig:ExpMethod}b). During the reduction process, ethanol and argon gases were introduced at flow rates of 2 sccm and 20 sccm, respectively. The process was conducted at $\rm 1200 ^\circ C$ for 1 h under a pressure of 40 Pa (0.3 Torr). The surface morphology of the samples was examined by scanning electron microscopy (SEM; Hitachi High-Tech, S-4800) at an acceleration voltage of 1.0 kV. Raman spectra of the samples were acquired using a Raman spectrometer (HORIBA Jovin Yvon, LabRAM HR800) with a 532 nm excitation.

\textbf{Nanoindentation measurements} The mechanical response of the rGO and rGO--CNT films under loading and unloading was examined via nanoindentation measurements\cite{butt2005force}. The nanoindentation was performed using an atomic force microscope (AFM; Bruker, NanoWizard 3) combined with an inverted confocal microscope (Nikon, Ti2-E). Cantilevers (MikroMasch, CSC-37-A/NoAl) with a spring constant of 1.0--1.5 N/m were employed. By pressing the AFM probe downward to contact the sample surface and then retracting it upward, the loading--unloading force--distance curves were obtained (Figure \ref{fig:ExpMethod}c). On each sample surface, force--distance curve measurements were performed at $\rm 32\times32$ points with a scanning area of $\rm 100\times100 \ \mu m$, thereby providing the in-plane distribution of the force-distance curves. The measurements were sequentially performed with increasing nanoindentation force from 15 nN to 300 nN, and the nanoindentation at 300 nN was repeated three times. For the calculation of dissipation energy from the force--distance curves, JPK data processing software (Bruker, version 8.1.16) was used.

\section{Results and Discussion}
\subsection{Stable configurations of Gr--CNT stacked structures in MD simulations}
\begin{figure*}[!htbp]
	\includegraphics[width=1\textwidth]{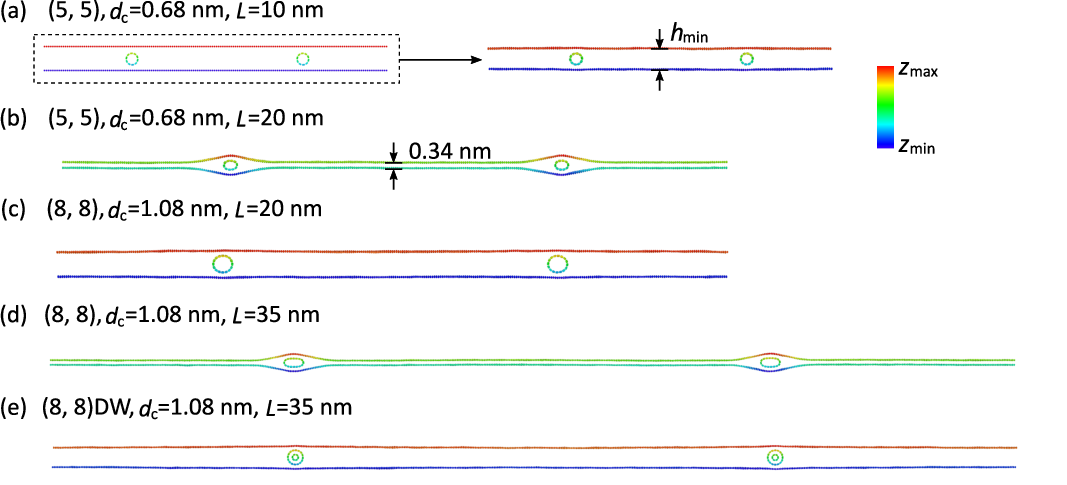}
	\centering
	\caption{Relaxed configurations obtained from MD simulations of the Gr--CNT stacked structure with different intertube distances $L$, CNT diameters $d_c$, and wall numbers. The color bar represents the z-coordinate of each atom, where $z_{\rm min}$ (blue) corresponds to the minimum z-coordinate and $z_{\rm max}$ (red) corresponds to the maximum z-coordinate. (a) Single-walled CNTs with a chiral index of (5, 5), a $d_c$ of 0.68 nm, and an $L$ of 10 nm. The dashed frame indicates the initial configuration. (b) Single-walled CNTs with a chiral index of (5, 5), a $d_c$ of 0.68 nm, and an $L$ of 20 nm. (c, d) Single-walled CNTs with a chiral index of (8, 8), a $d_c$ of 1.08 nm, and $L$ values of (c) 20 nm and (d) 35 nm. (e) Double-walled (DW) CNT with an outer-wall chiral index of (8, 8), a $d_c$ of 1.08 nm, and an $L$ of 35 nm.}
	\label{fig:DouFreDeform}
\end{figure*}

In Gr--CNT hybrid systems, it is desirable for Gr to exist as fully isolated single layers rather than directly-stacked multilayers because a higher fraction of single layers increases the accessible surface area and enhances properties such as adsorption capacity and reaction efficiency. The mixing ratio of CNTs to Gr in the hybrid system strongly influences the microstructure, determining whether the Gr layers can be adequately separated. To clarify this relationship, we performed MD simulations to examine the effect of intertube distance $L$ on the separation of Gr layers, where a shorter $L$ corresponds to a higher proportion of CNTs in the hybrid system. The effects of the CNT diameter $d_c$ and wall number were also examined. Starting from a Gr--CNT stacked structure in which the two Gr layers were initially separated, the systems were relaxed. The relaxed configurations obtained under various structural parameters are shown in Figure \ref{fig:DouFreDeform}. In the first case of single-walled CNTs with $d_c$ = 0.68 nm and $L$ = 10 nm  (Figure \ref{fig:DouFreDeform}a), the Gr layers remain separated by the CNTs, with slight deformation and a reduction of the interlayer distance relative to the initial configuration. When $L$ is increased to 20 nm while keeping the $d_c$ value at 0.68 nm (Figure \ref{fig:DouFreDeform}b), the Gr layers adhere to each other via vdW interactions. The interlayer distance of the two Gr layers is 0.34 nm, which is close to that of carbon layers in graphite. When $L$ is fixed and $d_c$ is increased to 1.08 nm (Figure \ref{fig:DouFreDeform}c), the Gr layers remain separated, similar to the first case. However, when $L$ is increased to 35 nm (Figure \ref{fig:DouFreDeform}d), the Gr layers adhere even for a $d_c$ of 1.08 nm. The vdW adhesion between the Gr layers exerts pressure on the CNTs, causing deformation of CNTs into elliptical cross sections. In Figure \ref{fig:DouFreDeform}e, the CNTs are replaced with double-walled CNTs. The outer wall is the same as that in Figures \ref{fig:DouFreDeform}c and d with a chiral index of (8, 8) and a $d_c$ of 1.08 nm, while the inner wall has a chiral index of (3, 3) and a $d_c$ of 0.41 nm. The double-walled CNTs maintain separation between the Gr layers, and the cross-sectional deformation is less pronounced than in the case of single-walled CNTs. 

\begin{figure*}[!htbp]
	\includegraphics[width=1\textwidth]{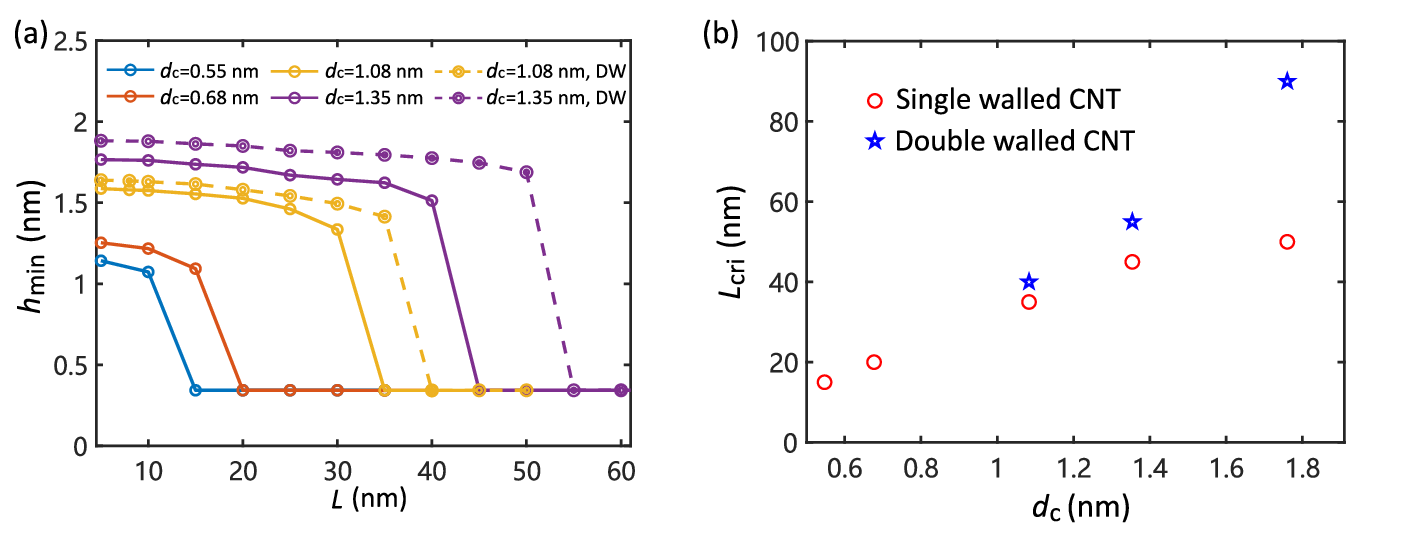}
	\centering
	\caption{Changes in the stable configurations of the Gr--CNT stacked structures obtained via MD simulations starting from the initial separation configuration. (a) Plots of minimum interlayer distance $h_{\rm min}$ against intertube distance $L$ for different CNT diameters $d_c$ and wall numbers. Single-walled CNTs are shown as open circles with solid lines: blue ($d_c$ = 0.55 nm), red ($d_c$ = 0.68 nm), yellow ($d_c$ = 1.08 nm), and purple ($d_c$ = 1.35 nm). Double-walled (DW) CNTs are shown as double circles with dashed lines: yellow ($d_c$ = 1.08 nm) and purple ($d_c$ = 1.35 nm). (b) Plots of critical intertube distance $L_{\rm cri}$ against $d_c$ for single-walled CNTs (red circles) and double-walled CNTs (blue stars).}
	\label{fig:DouFreLcri}
\end{figure*}

Based on the final interlayer distance, the stable configurations of the hybrid structure are classified into two distinct types, i.e., separation (Figures \ref{fig:DouFreDeform}a, c, and e) and adhesion (Figures \ref{fig:DouFreDeform}b and d). To distinguish both configurations, the minimum interlayer distance $h_{\rm min}$ in the relaxed state was calculated. To quantitatively understand the effect of CNT parameters on these configurations, additional simulations were conducted by varying $L$, $d_c$, and wall numbers. As shown in Figure \ref{fig:DouFreLcri}a, for the same $d_c$, $h_{\rm min}$ approaches the initial spacing $H$ when $L$ is small, indicating that the Gr layers remain fully separated. Owing to the relation $H=d_c+2\sigma$, $h_{\rm min}$ in the separation configuration increases with $d_c$. As $L$ increases and exceeds a critical value ($L_{\rm cri}$), $h_{\rm min}$ sharply decreases to 0.34 nm, indicating the occurrence of interlayer adhesion. A large $d_c$ increases the $L_{\rm cri}$ (Figure \ref{fig:DouFreLcri}b). Since the vdW adhesion force decreases with increasing distance, a larger $H$ weakens the interlayer attraction and prevents the interlayer adhesion. The increasing trend of $L_{\rm cri}$ with $d_c$ becomes less pronounced as $d_c$ increases. For single-walled CNTs with small $d_c$ values, $L_{\rm cri}$ increases from 20 nm to 35 nm with increasing $d_c$ from 0.68 nm to 1.08 nm. However, for those with larger $d_c$ values, $L_{\rm cri}$ only increases from 45 nm to 50 nm upon increasing $d_c$ from 1.35 nm to 1.76 nm. This tendency can be attributed to the radial collapse of single-walled CNTs, which may occur either spontaneously or under pressure \cite{meng_theoretical_2019,magnin_collapse_2021,heo_microstructural_2024,galafassi_reversible_2024}. The collapsed cross-section is energetically stable for single-walled CNTs when $d_c$ exceeds a critical value \cite{meng_theoretical_2019,magnin_collapse_2021,heo_microstructural_2024,galafassi_reversible_2024}. The critical $d_c$ for spontaneous collapse has been predicted to be 2--5 nm using different models and methods \cite{hasegawa_radial_2006,lu_radial_2011,lu_erratum_2011}. The pressure-induced collapse has also been examined \cite{tangney_structural_2005}. Since the critical pressure of the collapse follows $P \propto 1/d_c^3$ \cite{sun_pressure-induced_2004}, less pressure is required to induce collapse in CNTs with larger $d_c$ values. In the Gr--CNT system, the adhesion force of Gr layers between two CNTs is balanced by the supporting force of the CNTs, thereby generating compressive pressure on the CNTs. An increase in $L$ leads to a greater total adhesion force and consequently a larger compressive force on the CNTs. For CNTs with larger $d_c$, less pressure is sufficient to cause collapse; thus, the influence of increasing $d_c$ on $L_{\rm cri}$ becomes weaker. Since a larger CNT wall number can resist collapse under higher critical pressure \cite{meng_theoretical_2019,magnin_collapse_2021}, double-walled CNTs further increase $L_{\rm cri}$ compared with single-walled CNTs having the same $d_c$ (dotted lines in Figure \ref{fig:DouFreLcri}a). The difference in $L_{\rm cri}$ between single- and double-walled CNTs is pronounced especially at larger $d_c$ (blue stars in Figure \ref{fig:DouFreLcri}b).

\begin{figure*}[!htbp]
	\includegraphics[width=0.8\textwidth]{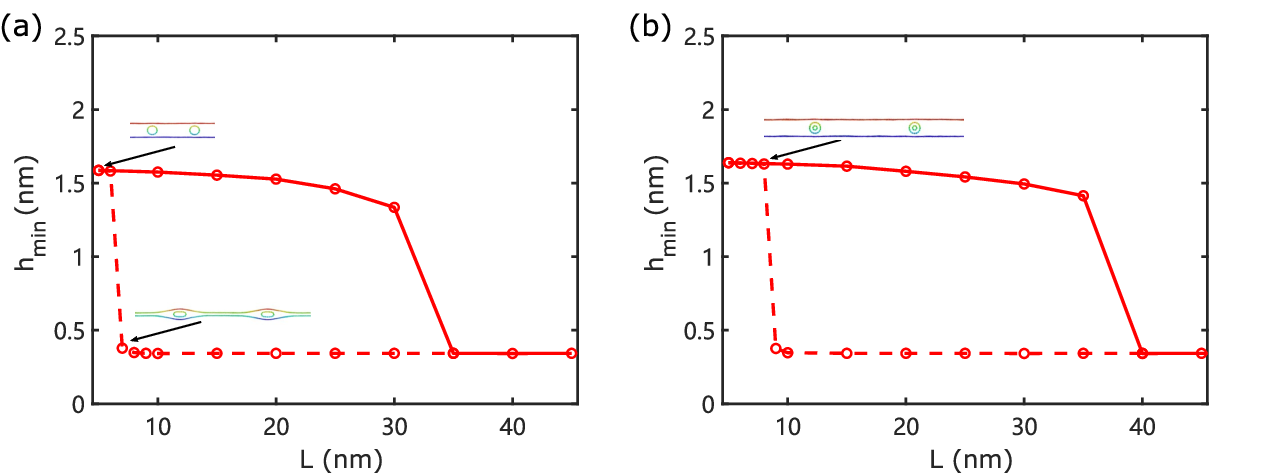}
 	\centering
	\caption{Bistability of Gr--CNT stacked structures revealed from MD simulations. Minimum interlayer distance $h_{\rm min}$ is plotted against intertube distance $L$ obtained using the initial separation configuration (solid lines) and the initial adhesion configuration (dotted lines). (a) Single-walled CNTs with a chiral index of (8, 8) and a CNT diameter $d_{c}$ of 1.08 nm. (b) Double-walled CNTs with an outer-wall chiral index of (8, 8) and a $d_{c}$ of 1.08 nm.}
	\label{fig:multistable}
\end{figure*}

In the simulations conducted above, separated Gr layers were selected as the initial configuration to examine whether the layers remained separated or adhered after relaxation. However, the resulting stable structure may depend on the initial configuration. Therefore, simulations starting from the initial adhesion configuration were also conducted. The adhesion structures were obtained by applying a uniform compressive force to each Gr atom of the original structure (Figure \ref{fig:MDsystem}c) until the Gr layers adhered. Then, the loading force was removed, and the adhesion structure was relaxed to reach a stable state. Figure \ref{fig:multistable}a shows the plots of $h_{\rm min}$ against $L$ for (8,8) single-walled CNTs starting from the initial adhesion configuration (dashed line) and from the initial separation configuration (solid line). The plot for the initial adhesion configuration shows the two stable states, separation and adhesion, depending on $L$. For shorter $L$ values, the Gr layers spontaneously separate after the external force is removed. Conversely, the adhesion structure remains stable after relaxation for larger $L$ values. The initial adhesion and separation configurations lead to different $L_{\rm cri}$ values of 7 nm and 35 nm, respectively. To distinguish them, the value obtained from the initial adhesion configuration is defined as $L_{\rm cri1}$, and that from the initial separation configuration as $L_{\rm cri2}$. Remarkably, when $L$ lies between $L_{\rm cri1}$ and $L_{\rm cri2}$, both separation and adhesion configurations are energetically stable, indicating the bistable nature of the system, in which the final structure depends on the initial state. Simulations with double-walled CNTs were also conducted, exhibiting the same bistable nature with a larger $L_{\rm cri1}$ of 9 nm (Figure \ref{fig:multistable}b).

\subsection{Energy analysis of Gr--CNT stacked structures}
Since the relaxation process follows the principle of minimum potential energy, the energy components were analyzed to clarify the physical origin of the bistability observed in the Gr--CNT stacked structures. The total potential energy $E_\text{pot}$ of the simulation system can be decomposed into three components as follows:
 \begin{equation}
E_\text{pot}=E_\text{Gr}+E_\text{CNT}+U_\text{LJ}.
\label{eqn:ch5a-Epot}
\end{equation}
Here, $E_\text{Gr}$ represents the internal energy of the two Gr layers described by the AIREBO potential. $E_\text{CNT}$ denotes the internal energy of the CNTs, including the AIREBO potential for the C--C bonds and the LJ energy within the CNT walls. $U_\text{LJ}$ corresponds to the LJ interaction, which can be further decomposed into the interaction between upper and lower Gr layers $U_\text{GrU-L}$, the interaction between the upper Gr layer and the CNTs $U_\text{GrU-CNT}$, and the interaction between the lower Gr layer and the CNTs $U_\text{GrL-CNT}$. Among these subcomponents of $U_\text{LJ}$, $U_\text{GrU-L}$ contributes the most to $U_\text{LJ}$ because the contact area between the Gr layers and the CNTs is small compared with the total Gr area.

For each component in Eq. \ref{eqn:ch5a-Epot}, the energy difference between the separation (Sep) and adhesion (Adh) configurations was calculated and expressed using a $\Delta$ operator (e.g., $\Delta E_\text{pot}=E_\text{pot, Adh}-E_\text{pot, Sep}$). Accordingly, the total potential energy difference $\Delta E_\text{pot}$ is given as follows:
\begin{equation}
\Delta E_\text{pot}= \Delta E_\text{Gr}+ \Delta E_\text{CNT}+\Delta U_\text{LJ}.
\label{eqn:ch5a-EpotDelat}
\end{equation}
In the adhesion configuration, the vdW potential energy is lower than that in the separation configuration ($\Delta U_\text{LJ}<0$) because the LJ energy decreases with decreasing interlayer distance $h$ when $h>0.34$ nm. The magnitude of this energy difference increases with increasing size of the Gr layers, i.e., $\Delta U_\text{LJ}\propto L$. Deformation of the Gr layers leads to bending of the C--C bonds, which in turn increases the internal energy ($\Delta E_\text{Gr}>0$). In the adhesion configuration, the pressure exerted by the Gr layers deforms the CNTs into an elliptical cross-section, thereby increasing the internal energy ($\Delta E_\text{CNT}>0$).

\begin{figure*}[!htbp]
	\includegraphics[width=0.95\textwidth]{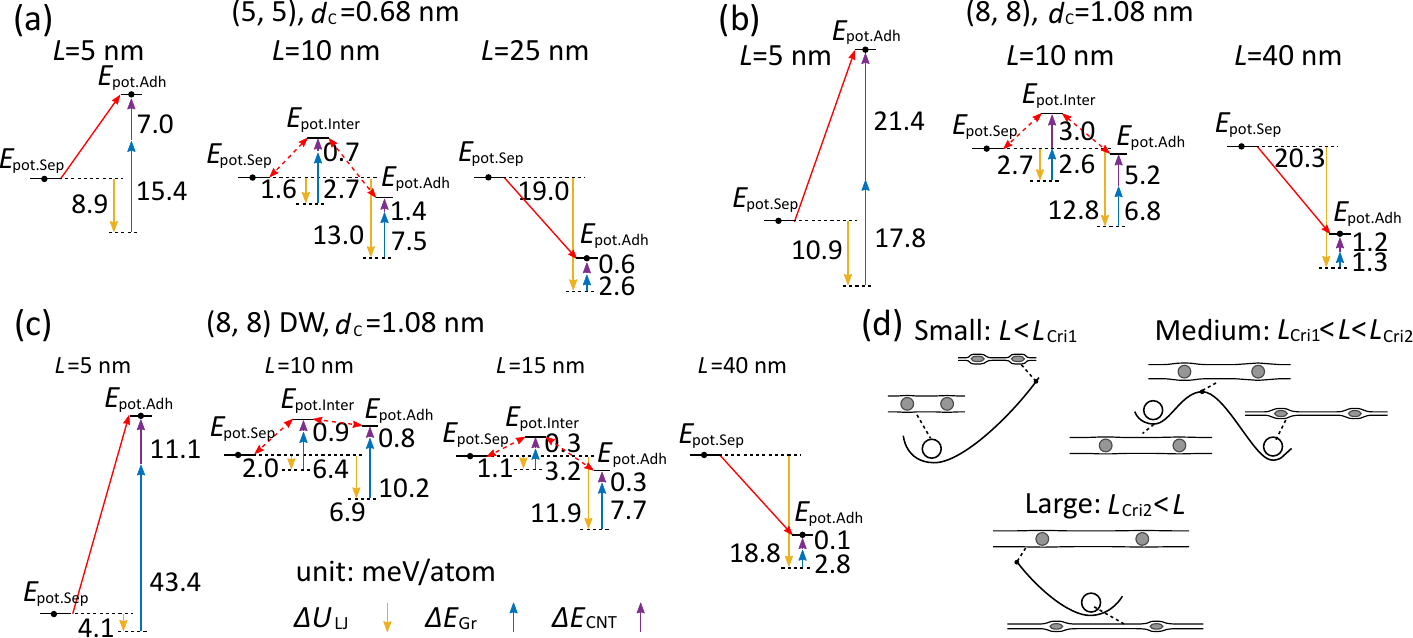}
 	\centering
	\caption{Energy component analysis of the Gr--CNT stacked structures via MD simulations. (a--c) Potential energy ($E_\text{pot}$) diagrams for the separation (Sep), intermediate (Inter), and adhesion (Adh) configurations, calculated for different intertube distance $L$, CNT diameter $d_c$, and wall number values. The numeral values indicate energy in meV/atom. The red arrows indicate the difference in total potential energy $\Delta E_\text{pot}$. The yellow arrows represent the difference in vdW energy by LJ potential $\Delta U_{\rm LJ}$. The blue and purple arrows represent the difference in internal energy of Gr $\Delta E_{\rm Gr}$ and CNT $\Delta E_{\rm CNT}$, respectively. (a) Single-walled CNTs with a chiral index of (5, 5), a $d_c$ of 0.68 nm, and $L$ values of 5, 10, and 25 nm. (b) Single-walled CNTs with a chiral index of (8, 8), a $d_c$ of 1.08 nm, and $L$ values of 5, 10, and 40 nm. (c) Double-walled CNTs with an outer-wall chiral index of (8, 8), a $d_c$ of 1.08 nm, and $L$ values of 5, 10, 15, and 40 nm. (d) Schematic energy landscapes of the separation and adhesion configurations for the three $L$ regimes: ``Small'', ``Medium'', and ``Large''.}
	\label{fig:ch5a-EnergyAna}
\end{figure*}

Figures \ref{fig:ch5a-EnergyAna}a--c compare the energy components for various $L$ values and CNT parameters, while the schematic energy landscapes depending on $L$ relative to $L_{\rm cri1}$ and $L_{\rm cri2}$ are shown in Figure \ref{fig:ch5a-EnergyAna}d. For $L<L_{\rm cri1}$ (e.g., $L=5$ nm in Figure \ref{fig:ch5a-EnergyAna}a), the increase in $\Delta E_\text{Gr}+\Delta E_\text{CNT}$ exceeds the decrease in $\Delta U_\text{LJ}$, making the separation configuration the most stable ($\Delta E_\text{pot}>0$), depicted as ``Small'' in Figure \ref{fig:ch5a-EnergyAna}d. As $L$ increases, the reduction in the vdW energy of the Gr layers upon adhesion becomes more pronounced ($\Delta U_\text{LJ} \propto L$), resulting in a lower $E_\text{pot}$ for the adhesion configuration ($\Delta E_\text{pot}<0$). Thus, for $L>L_{\rm cri2}$ (e.g., $L=25$ nm in Figure \ref{fig:ch5a-EnergyAna}a), the adhesion configuration becomes the most stable state regardless of the initial configuration (``Large'' in Figure \ref{fig:ch5a-EnergyAna}d). At intermediate $L$ values ($L_{\rm cri1}<L<L_{\rm cri2}$; e.g., $L=10$ nm in Figure \ref{fig:ch5a-EnergyAna}a), the system exhibits bistability with two locally stable states corresponding to the adhesion and separation configurations, owing to the balance between vdW attraction and elastic deformation energy. This bistability is associated with a potential barrier that prevents spontaneous transitions between the two configurations (``Medium'' in Figure \ref{fig:ch5a-EnergyAna}d). During the transition from separation to adhesion, the approach of the Gr layers decreases $U_\text{LJ}$, while the bending of the Gr layers increases both $E_\text{Gr}$ and $E_\text{CNT}$. As a result, $E_\text{pot}$ initially increases from the separation state to the intermediate (Inter) state ($E_\text{pot,Sep}<E_\text{pot,Inter}$), indicating that the transition does not occur spontaneously. The reverse transition from adhesion to separation is also non-spontaneous ($E_\text{pot,Adh}<E_\text{pot,Inter}$) because of the strong vdW adhesion between the Gr layers.

The deformations of Gr layers and CNTs in the adhesion configuration increase their respective internal energies ($\Delta E_\text{Gr}$ > 0 and $\Delta E_\text{CNT}$ > 0). The increase in $\Delta E_\text{CNT}$ is more pronounced for CNTs with larger diameters (Figures \ref{fig:ch5a-EnergyAna}a and b). For CNTs with larger $d_c$, a larger $\Delta U_\text{Gr}$ (i.e., larger $L$) is required to overcome the energy barrier, thereby increasing $L_{\rm cri}$. With respect to the wall number, double-walled CNTs are more resistant to collapse under the pressure exerted by the Gr layers than single-walled CNTs. In the adhesion configuration, the Gr layers exhibit greater bending to conform to the circular cross-section of double-walled CNTs. The energy barrier mainly arises from $\Delta E_\text{Gr}$, whereas the contribution of $\Delta E_\text{CNT}$ is minor (Figure \ref{fig:ch5a-EnergyAna}c). Consequently, the sum of $\Delta E_\text{Gr}$ and $\Delta E_\text{CNT}$ for double-walled CNTs is larger than that for single-walled CNTs, leading to a higher $L_{\rm cri}$.

\subsection{Loading--unloading behavior and hysteresis of Gr--CNT stacked structures}
The Gr--CNT hybrid structure is known to exhibit superelastic behavior with ultraflexibility and hysteretic loading--unloading response \cite{kim_graphene_2012,li_fabrication_2017,afroze_hierarchical_2021}. To further examine this property, the loading--unloading behavior of the Gr--CNT structure was simulated under pressure. The structures that remained in the separation configuration after relaxation were employed for these simulations. A uniform vertical force was applied to all atoms in the Gr layers to impose the pressure (Figure \ref{fig:MDsystem}c). The deformation was quantified using the volume strain $\varepsilon_V$, which is defined as $\varepsilon_V = (V_0 - V)/V_0$, where $V_0$ is the initial interlayer volume and $V$ is the deformed volume during loading. 

\begin{figure*}[!htbp]
	\includegraphics[width=0.8\textwidth]{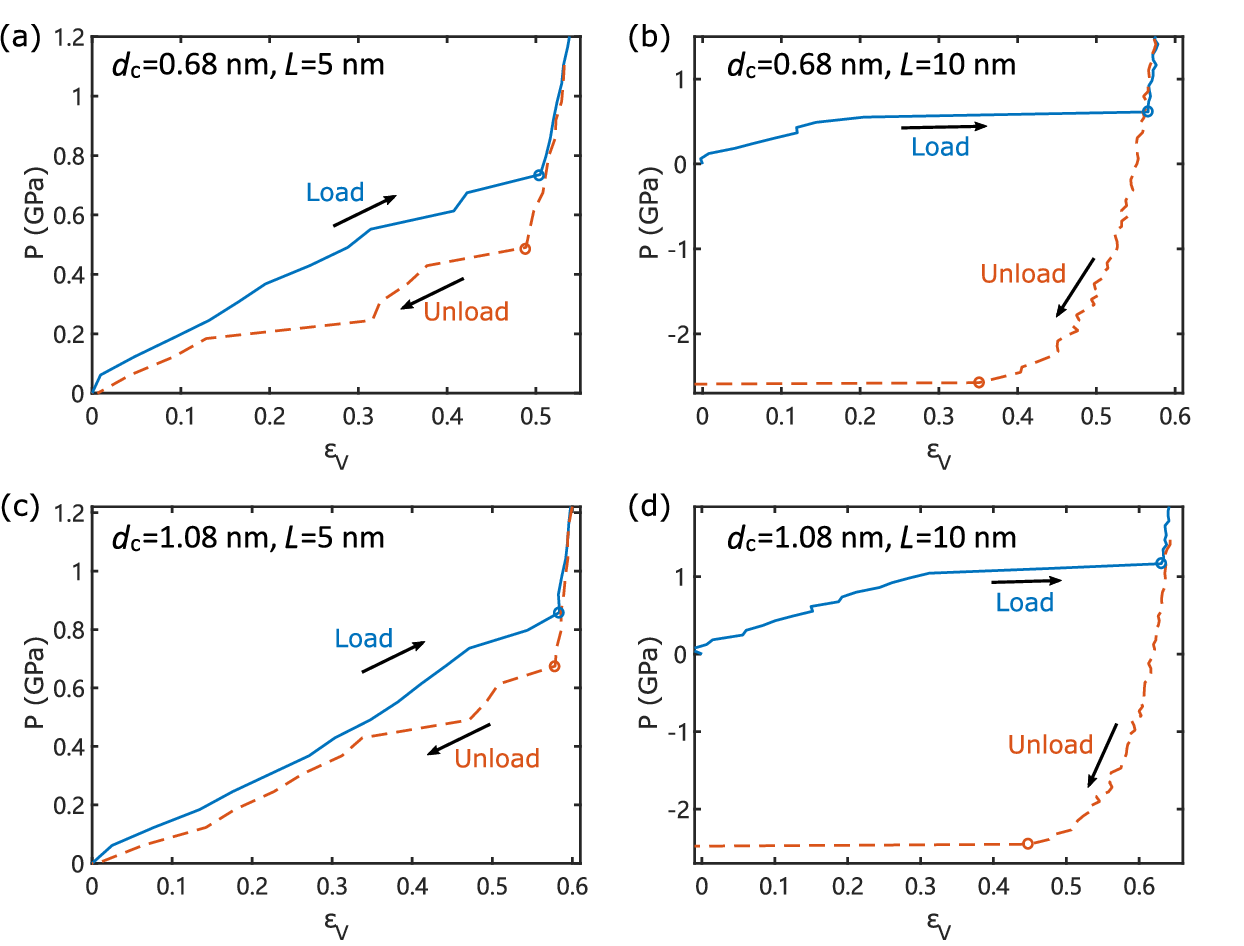}
 	\centering
	\caption{Loading--unloading response of Gr--CNT stacked structures obtained via MD simulations, showing pressure $P$ as a function of volume strain $\varepsilon_V$. Loading and unloading curves are represented by solid blue lines and dashed red lines, respectively. The critical points of the Gr separation are indicated by hollow circles. (a) $d_c = 0.68$ nm, $L = 5$ nm. (b) $d_c = 0.68$ nm, $L = 10$ nm. (c) $d_c = 1.08$ nm, $L = 5$ nm. (d) $d_c = 1.08$ nm, $L = 10$ nm. All CNTs are single-walled.}
	\label{fig:addforce}
\end{figure*}

The loading--unloading curves, namely the relationship between $\varepsilon_V$ and pressure $P$, are plotted for the structure with $d_c=0.68$ nm and $L=5$ nm in Figure \ref{fig:addforce}a. The interlayer distance decreases gradually as $P$ increases from 0, which leads to a decrease in $V$ and an increase in $\varepsilon_V$. When the interlayer distance approaches 0.34 nm, the Gr layers adhere together. A further increase in $P$ leads to only a small additional decrease in the interlayer distance because of the repulsive term in the vdW force (Eq. \ref{eqn:LJ12_6}). Then, the pressure is unloaded by decreasing the force on each Gr atom. Upon unloading, the interlayer distance increases as the Gr layers separate. The critical force required for separation during unloading is smaller than that during loading (hollow circles in Figure \ref{fig:addforce}a). This behavior indicates a clear hysteresis, as the loading and unloading curves do not overlap. When $L$ increases to 10 nm while keeping the $d_c$ at 0.68 nm (Figure \ref{fig:addforce}b), the critical loading force becomes smaller than that for $L=5$ nm. Moreover, the unloading curve exhibits a qualitatively different behavior from the previous case. The Gr layers remain adhered even when the applied force is reduced to zero, and a negative (peeling) force is required to separate the Gr layers. If the negative pressure is released after the layers are fully separated ($\varepsilon_V=0$), the structure remains in the separation configuration. This cycle can also be regarded as exhibiting hysteresis, although the layer separation does not occur spontaneously during unloading. When $d_c$ is large (1.08 nm, Figure \ref{fig:addforce}c), a hysteretic behavior in positive $P$ is observed, and the critical force is higher than that in Figure \ref{fig:addforce}a. A similar peeling force is also required when $d_c$ is 1.08 nm and $L$ is 10 nm (Figure \ref{fig:addforce}d), indicating that the two different hysteretic behaviors occur depending on $L$ even with larger $d_c$.

These loading--unloading results can be interpreted in terms of the structural stability discussed above. For small intertube distances ($L < L_{\rm cri1}$), the separation configuration is energetically favored. However, the clamping effect of the approaching Gr layers imposes additional pressure on the CNTs, which induces the intrinsic hysteretic deformation of the CNTs \cite{meng_theoretical_2019,tangney_structural_2005}. This is reflected as the measurable hysteresis of the entire Gr--CNT system (Figures \ref{fig:addforce}a and c). For intermediate $L$ values ($L_{\rm cri1} < L < L_{\rm cri2}$), the system resides within the bistable regime, where both separation and adhesion configurations are locally stable. In this case, an external peeling force is necessary to overcome the potential barrier associated with vdW attraction and the elastic deformation of the Gr layers and CNTs, leading to pronounced hysteresis (Figures \ref{fig:addforce}b and d). Larger CNT diameters increase the contribution of the elastic deformation energy, thereby increasing the critical force required for layer separation. Overall, the hysteretic responses directly reflect the interplay between vdW attraction and elastic deformation, in agreement with the energy analysis presented in the previous section. Such multistability has also been exploited in architected macroscopic materials to store and release elastic strain energy\cite{shan2015multistable}. The present results indicate that an analogous mechanism emerges even in microstructures constructed from nanomaterials, suggesting that multistability--based functionalities can be extended down to nanoscale building blocks.

\subsection{Experimental observation of hysteresis in Gr--CNT hybrid films}
\begin{figure*}[!htbp]
	\includegraphics[width=0.95\textwidth]{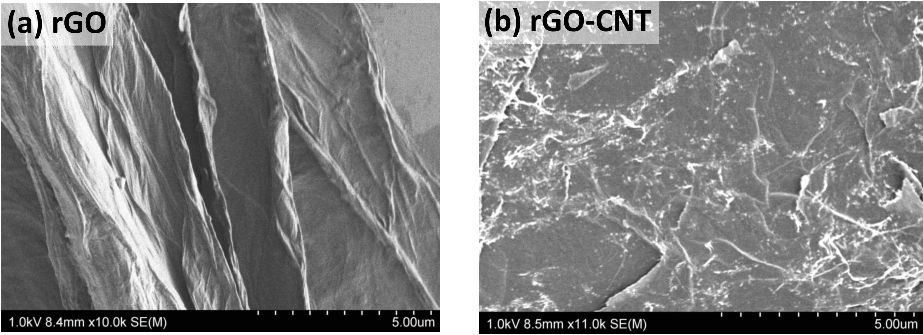} 
	\centering
	\caption{SEM images of the samples used for nanoindentation experiments: (a) rGO film and (b) rGO--CNT hybrid film.}
	\label{fig:Ch5-SEM}
\end{figure*}

To experimentally verify the mechanical hysteresis observed in the MD simulations, the mechanical behavior of the Gr--CNT hybrid system was characterized using AFM-based nanoindentation. Submillimeter-thick rGO--CNT hybrid films composed of randomly oriented rGO sheets and CNTs were employed as macroscopic assemblies of the structure used in the MD simulations. In the MD simulations, a simple unit consisting of two Gr layers sandwiching CNTs was examined as the fundamental structural element. The hybrid film in the experiment can therefore be regarded as a stacked network composed of a large number of such unit structures. In addition to pure rGO films as a reference, rGO--CNT hybrid films were prepared using vacuum filtration followed by high-temperature reduction. Figure \ref{fig:Ch5-SEM} presents SEM images of the surfaces of the rGO and rGO--CNT films. The wrinkles on the surface of the rGO film were caused by solvent evaporation and surface tension during the transfer of the film from acetone to the substrate. Meanwhile, the rGO--CNT film surface exhibits fibrous structures, indicating the presence of CNTs embedded within the film. The Raman spectra of the samples indicate relatively high intensity ratios of the G band to the D band compared with typical rGO obtained by low-temperature chemical reduction\cite{stankovich2007synthesis}, supporting effective structural restoration by the high-temperature treatment (Figure S1).

\begin{figure*}[!htbp]
	\includegraphics[width=0.95\textwidth]{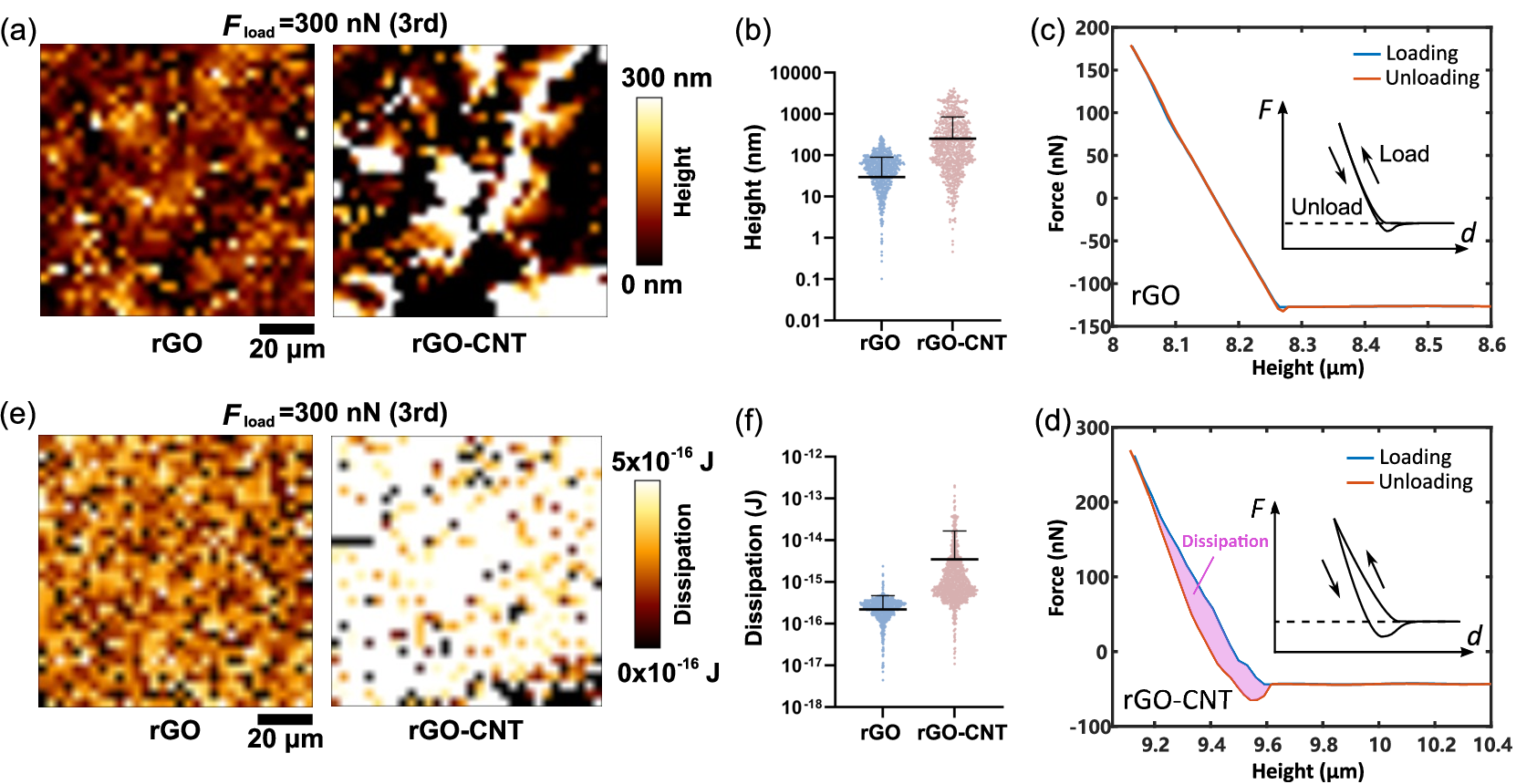}
	\centering
	\caption{Experimental verification of hysteresis in the Gr--CNT hybrid system using nanoindentation. (a) AFM surface images of rGO and rGO--CNT films obtained in the third measurement under a loading force of 300 nN. (b) Histogram of the height distribution for the rGO and rGO--CNT films. (c, d) Force--distance curves from loading--unloading processes for the (c) rGO and (d) rGO--CNT films. Loading and unloading curves are shown in blue and red, respectively. (e) Dissipation energy mapping for the rGO and rGO--CNT films. (f) Histogram of dissipation energy distribution for the rGO and rGO--CNT films.}
	\label{fig:Ch5-AFMNanoindent}
\end{figure*}

Figure \ref{fig:Ch5-AFMNanoindent}a shows the AFM images displaying the surface morphology of the rGO and rGO--CNT films. Compared with the relatively smooth surface of the rGO film, certain regions in the rGO--CNT hybrid film exhibit larger height variations, reflecting the separation between rGO layers induced by the insertion of CNTs. This uneven distribution originates from CNT aggregation: in regions with higher CNT density, the rGO layers are more separated. Statistical analysis of multiple height points (Figure \ref{fig:Ch5-AFMNanoindent}b) also indicates that the addition of CNTs leads to greater interlayer separation of the rGO sheets. Height mapping conducted before each loading process (Figure S2) confirms that the height increase caused by the CNTs is independent of the indentation force.

Pressure loading--unloading curves of the samples were obtained using AFM-based nanoindentation. Figures \ref{fig:Ch5-AFMNanoindent}c and d show representative force--distance curves measured at a single point for the rGO and rGO--CNT films, respectively. For the rGO film, the loading and unloading curves nearly overlap, indicating negligible hysteresis, whereas pronounced hysteresis is observed for the rGO--CNT film. This demonstrates that the introduction of CNTs significantly alters the mechanical response of the Gr assemblies.

The observed hysteresis in the rGO--CNT film closely resembles the behavior revealed by the MD simulations. Note that the experimental force--distance curves (Figure \ref{fig:Ch5-AFMNanoindent}d) correspond to left--right inversions of the MD curves (Figure \ref{fig:addforce}), owing to the difference between the probe height (distance) and volume strain used for the $x$--axes. During the loading process, a larger force is required to compress the structures, whereas the force applied during unloading is smaller. As a result, the mechanical work input during loading is larger than the energy released during unloading. The difference in external work applied during the loading and unloading processes demonstrates the occurrence of energy dissipation. Similar macroscopic dissipation has been reported for rGO--CNT foams used in strain sensors and energy-absorbing structures \cite{kim_graphene_2012,afroze_hierarchical_2021}. Comparing the results of the rGO and rGO--CNT films reveals that the introduction of CNTs induces this dissipation behavior. In the actual samples, CNTs are not uniformly distributed in the film and exhibit local variations in the density. As a result, different regions within the rGO--CNT film correspond to different effective intertube distances: some regions satisfy $L < L_{\rm cri1}$, while others fall within the intermediate regime $L_{\rm cri1} < L < L_{\rm cri2}$. Since AFM indentation only imposes compressive loading globally, the dominant contribution to the experimental hysteresis arises from the response shown in Figure \ref{fig:addforce}a. Nevertheless, deformation of surrounding units can locally create tensile-like forces on certain units, which correspond to negative pressure in the MD simulations. These local tensile forces may allow a minor contribution from the response shown in Figure \ref{fig:addforce}b. The macroscopic mechanical behavior of the film arises from the superposition of the responses from Gr--CNT structural units subjected to different local pressure states.

The area enclosed by the loading and unloading curves in the AFM measurements is defined as the dissipation energy (Figure \ref{fig:Ch5-AFMNanoindent}d). Figure \ref{fig:Ch5-AFMNanoindent}e shows the spatial distribution of the dissipation energy obtained from multiple-point measurements across the sample surface. The histogram in Figure \ref{fig:Ch5-AFMNanoindent}f presents the statistical distribution of the dissipation energy at various observation points for the two films. Higher dissipation energy is observed at multiple points of the rGO--CNT film compared with the rGO film, which arises from local variation in CNT density. Across all tested forces (15 nN, 100 nN, and 300 nN; Figure S3), the rGO--CNT film exhibits consistently higher dissipation energy than the rGO film. These results confirm that the insertion of CNTs is responsible for the observed energy dissipation.

\section{Conclusion}
In this study, we investigated the microstructural stability and mechanical response of Gr--CNT stacked systems. Two distinct stable states corresponding to separation and adhesion configurations were identified through MD simulations. The transition between the two configurations depends on the CNT parameters: a shorter intertube distance, a larger diameter, and a larger wall number favor the separation of Gr layers. Under certain conditions, the system exhibits bistability, where both separation and adhesion configurations are energetically stable and the final structure depends on the initial state. The energy analysis reveals that the bistability originates from the competition between interlayer vdW attraction and elastic deformation energies of Gr and CNTs. The loading--unloading curves obtained via MD simulations exhibit hysteresis behavior stemming from the energetic stability. In addition, rGO--CNT hybrid films were experimentally fabricated as macroscopic assemblies of the simulated structures. AFM-based nanoindentation measurements reveal that the rGO--CNT films exhibit clear hysteresis and higher dissipation energy compared with pure rGO, in good agreement with the simulation results. The insertion of CNTs is considered to induce local separation of the rGO layers and to enhance the energy dissipation during loading cycles. These findings provide microscopic insight into the structure--response relationship of Gr--CNT hybrid systems and a direct connection between atomistic mechanisms and macroscopic mechanical behaviors. This knowledge will guide the design of Gr-based hybrid structure applications, not only in devices exploiting mechanical responses but also in systems where structural changes passively influence performance. Moreover, since the present hybrid systems are classified as mixed-dimensional vdW heterostructures\cite{jariwala2017mixed} composed of two-dimensional Gr and one-dimensional CNTs, the obtained insights may also be applicable to other combinations of various nanomaterials, including nanosheets\cite{novoselov2005two, geim2013van}, nanotubes\cite{tenne1992polyhedral,chopra1995boron, xiang2020one}, and nanoscrolls\cite{viculis2003chemical, zhao2021high}.

\begin{acknowledgement}


This research was supported by JSPS KAKENHI (JP19H04545, JP21H01763, JP24K01297, JP25K01623, and JP25K22205), JST SPRING (JPMJSP2138), JST PRESTO (JPMJPR24H1), JST FOREST (JPMJFR205N), and SAKIGAKE Club OU Ecosystem Support Program. The MD simulations of this research work utilized the computing resources of the Institute of Solid State Physics (ISSP) Supercomputer Center, The University of Tokyo. The SEM measurements were performed at the Research Center for Ultra-High Voltage Electron Microscopy, The University of Osaka. 

\end{acknowledgement}

\bibliography{ref}

\clearpage
\section*{Supporting Information}

\setcounter{figure}{0}
\setcounter{table}{0}
\setcounter{equation}{0}
\renewcommand{\thefigure}{S\arabic{figure}}
\renewcommand{\thetable}{S\arabic{table}}
\renewcommand{\theequation}{S\arabic{equation}}

\begin{figure*}[!htbp]
	\includegraphics[width=0.7\textwidth]{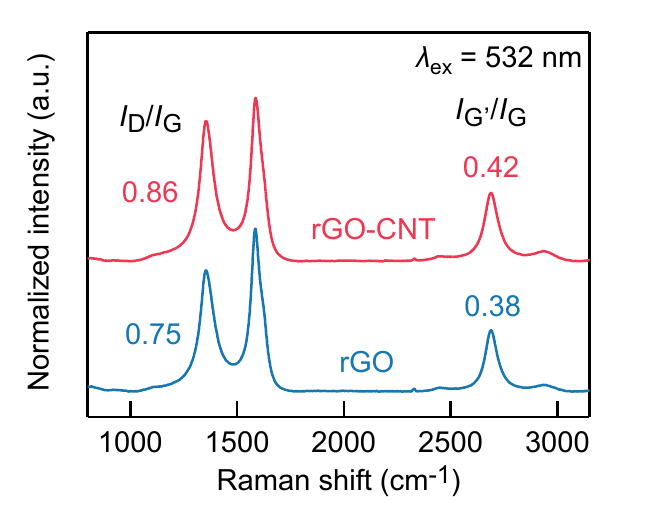}
	\centering
	\caption{Raman spectra of the rGO and rGO--CNT films acquired at an excitation laser wavelength of 532 nm. The D band to G band intensity ratios $I_{\rm D}/I_{\rm G}$ are 0.75 and 0.86 for the rGO and rGO--CNT films, respectively. The G' band to G band intensity ratios $I_{\rm G'}/I_{\rm G}$ are 0.38 and 0.42 for the rGO and rGO--CNT films, respectively. }
	\label{fig:supplRaman}
\end{figure*}

\newpage

\begin{figure*}[!htbp]
	\includegraphics[width=0.95\textwidth]{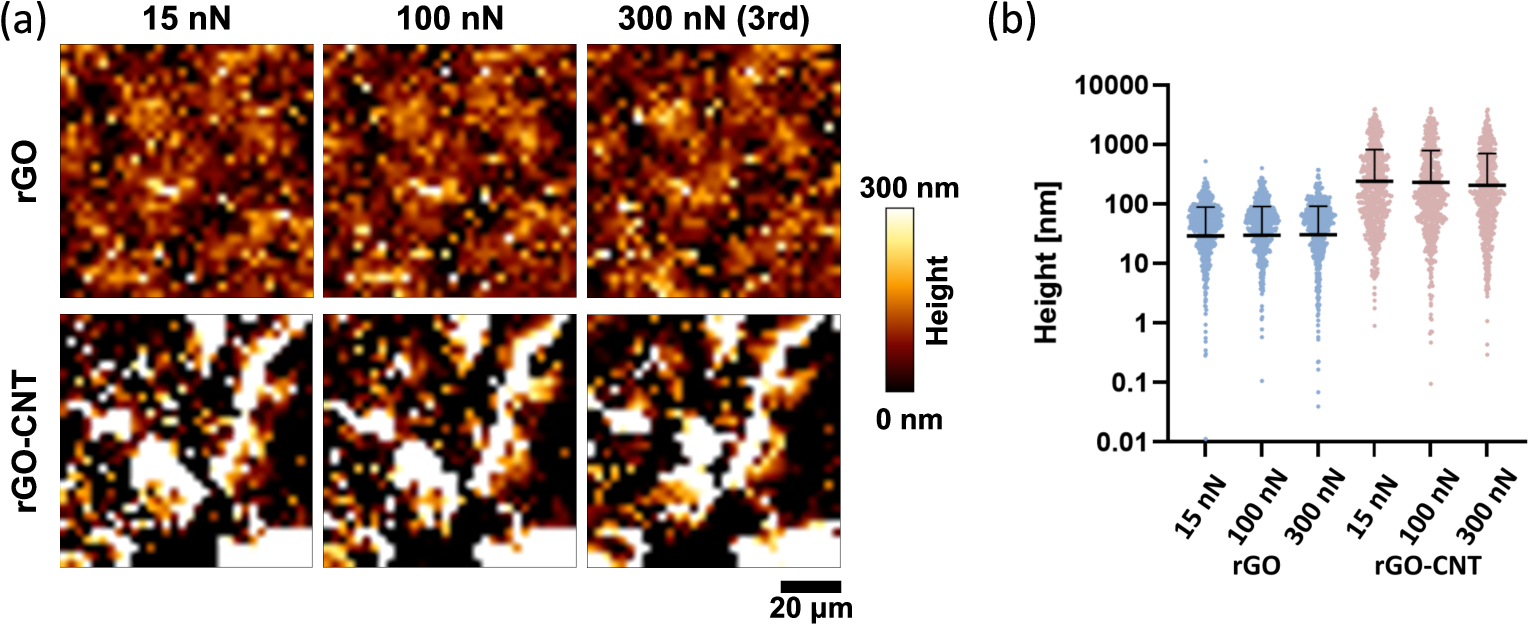}
	\centering
	\caption{(a) AFM surface images of the rGO and rGO--CNT films with different indentation forces (15 nN, 100 nN, and the third time of 300 nN). (b) Histogram of the height distribution for the rGO and rGO--CNT films with different indentation forces.}
	\label{fig:supplHeight}
\end{figure*}

\begin{figure*}[!htbp]
	\includegraphics[width=0.95\textwidth]{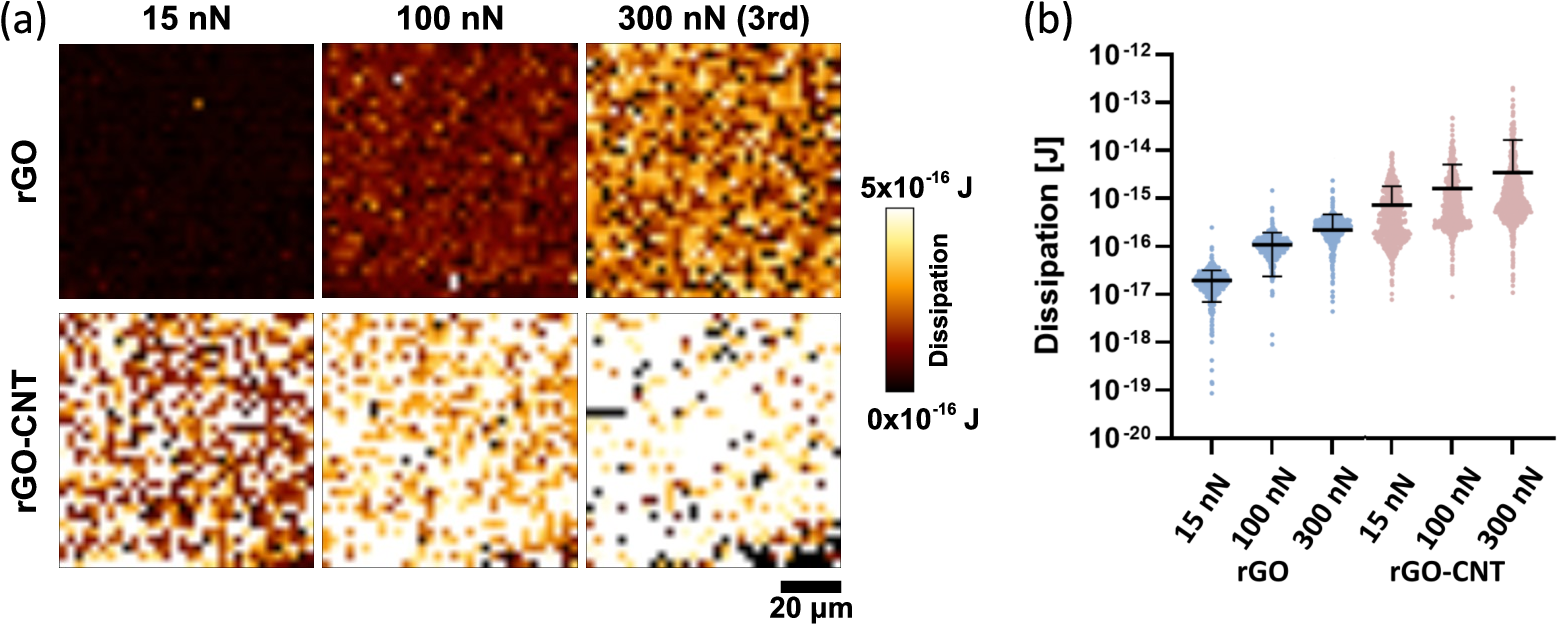}
	\centering
	\caption{(a) Dissipation energy mapping for the rGO and rGO--CNT films with different indentation forces (15 nN, 100 nN, and the third time of 300 nN) (b) Histogram of dissipation energy distribution for the rGO and rGO--CNT films with different indentation forces.}
	\label{fig:supplDissip}
\end{figure*}

\end{document}